\documentclass[a4paper]{article}

\usepackage{INTERSPEECH2021}
\usepackage{graphicx}
\usepackage{multirow}
\usepackage{subfigure}

\title{Deep Interaction between Masking and Mapping Targets for Single-Channel Speech Enhancement}
\name{Lu Zhang, Mingjiang Wang, Zehua Zhang, Xuyi Zhuang}
\address{
  Department of Electronics and Information Engineering,\\
  Harbin Institute of Technology, Shenzhen, China, 518000}
\email{\{18B952047,19S052011,19S052014\}@stu.hit.edu.cn, mjwang@hit.edu.cn}

\begin{document}

\maketitle
\begin{abstract}
  The most recent deep neural network (DNN) models exhibit impressive denoising performance in the time-frequency (T-F) magnitude domain. However, the phase is also a critical component of the speech signal that is easily overlooked. In this paper, we propose a multi-branch dilated convolutional network (DCN) to simultaneously enhance the magnitude and phase of noisy speech. A causal and robust monaural speech enhancement system is achieved based on the multi-objective learning framework of the complex spectrum and the ideal ratio mask (IRM) targets. In the process of joint learning, the intermediate estimation of IRM targets is used as a way of generating feature attention factors to realize the information interaction between the two targets. Moreover, the proposed multi-scale dilated convolution enables the DCN model to have a more efficient temporal modeling capability. Experimental results show that compared with other state-of-the-art models, this model achieves better speech quality and intelligibility with less computation.
\end{abstract}
\noindent\textbf{Index Terms}: complex spectrum, dilated convolution, multi-objective, multi-scale, speech enhancement

\section{Introduction}

Speech enhancement is a key front-end processing module, which is widely used in many speech-related products to extract the high-quality target speech from the noisy signals in adverse acoustic environments. According to the needs of different application scenarios for enhanced speech, speech enhancement can be divided into two categories, one is for humans and the other is for machines. For the needs of machines, such as speech recognition, the enhanced speech needs to reduce noise interference while minimizing speech distortion, thereby improving the accuracy and robustness in noisy environments. But for hearing-aids, voice calls, and other human hearing applications, more attention must be paid to improving the quality and intelligibility of the speech, and the time-delay must be strictly controlled.

Early research on monaural speech enhancement mainly paid more attention to the derivation of speech spectrum estimators, such as Wiener filtering \cite{scalart1996speech} and statistical-based estimators \cite{ephraim1985speech,cohen2001speech}. Those methods rely heavily on the tracking accuracy of the noise spectrum and often fail to handle the non-stationary noises. Recent advances in deep neural network (DNN)-based methods \cite{xu2014regression,kumar2016speech} have shown its powerful noise reduction ability in complex noise environments. This is due to the superior nonlinear modeling capabilities of the DNN models in time-frequency (TF) speech representations, like ideal ratio mask (IRM) or log power spectrum (LPS).

Although mapping or masking target modeling \cite{park2016fully,tan2018gated} in the TF magnitude domain has achieved remarkable results, noisy phase distortion needs to be solved for better speech quality and intelligibility. However, the phase spectrogram seems randomly distributed and unstructured \cite{gerkmann2015phase}, which is difficult to be processed directly. Therefore, some indirect solutions, such as complex spectral masking \cite{williamson2015complex,wang2019masking,hu2020dccrn}, complex spectral mapping \cite{li2020two,tan2019learning}, and waveform mapping \cite{pascual2017segan,fu2018end,pandey2019tcnn}, take phase information into account in DNN modeling and achieve further performance improvements. Since noise and speech signals are more easily distinguished in the TF domain, our work focuses on modeling in the TF complex domain to make full use of this important prior knowledge. In recent studies, although the convolutional recurrent network (CRN) \cite{hu2020dccrn,tan2019learning} structure has proved its model superiority in complex domain modeling, its training process is very time-consuming due to the adoption of recurrent neural networks. As an alternative scheme for temporal modeling, fully convolutional models \cite{li2020two,zhang2020multi} using dilated convolution show obvious advantages in both training and inference efficiency.

In this paper, we propose a novel phase-aware dilated convolutional network (DCN) model, named ‘PhaseDCN’, to achieve an efficient and robust speech enhancement system suitable for human hearing applications. A recent study \cite{li2020recursive} proves that incorporating dynamic attention in each frequency point to distinguish the noise or speech dominant components can result in better speech quality and less residual noises. Inspired by this, we integrate the masking target ideal ratio mask (IRM) into the complex spectrum reconstruction process for the first time and realize a DCN-based dual-path interactive learning framework. The generated frequency attention factors in the auxiliary path (IRM path) help to better recover the complex spectrum of the main path. The proposed PhaseDCN model not only reduces the phase distortion, but also realizes the complementary advantages of mapping and masking targets. While targeting the low-latency requirement, only causal dilated convolutions are applied and the proposed multi-scale encoding method makes our model more lightweight in terms of model size and computation complexity.

The rest of this paper is structured as follows. The proposed PhaseDCN model is described in Section 2. The experimental setup and results are presented in Section 3. Finally, Section 4 concludes this paper.

\section{Proposed Speech Denoising System}

\subsection{Feature extraction with dilated convolutions}

\begin{figure*}[t]
\centering
	\includegraphics[scale=0.5]{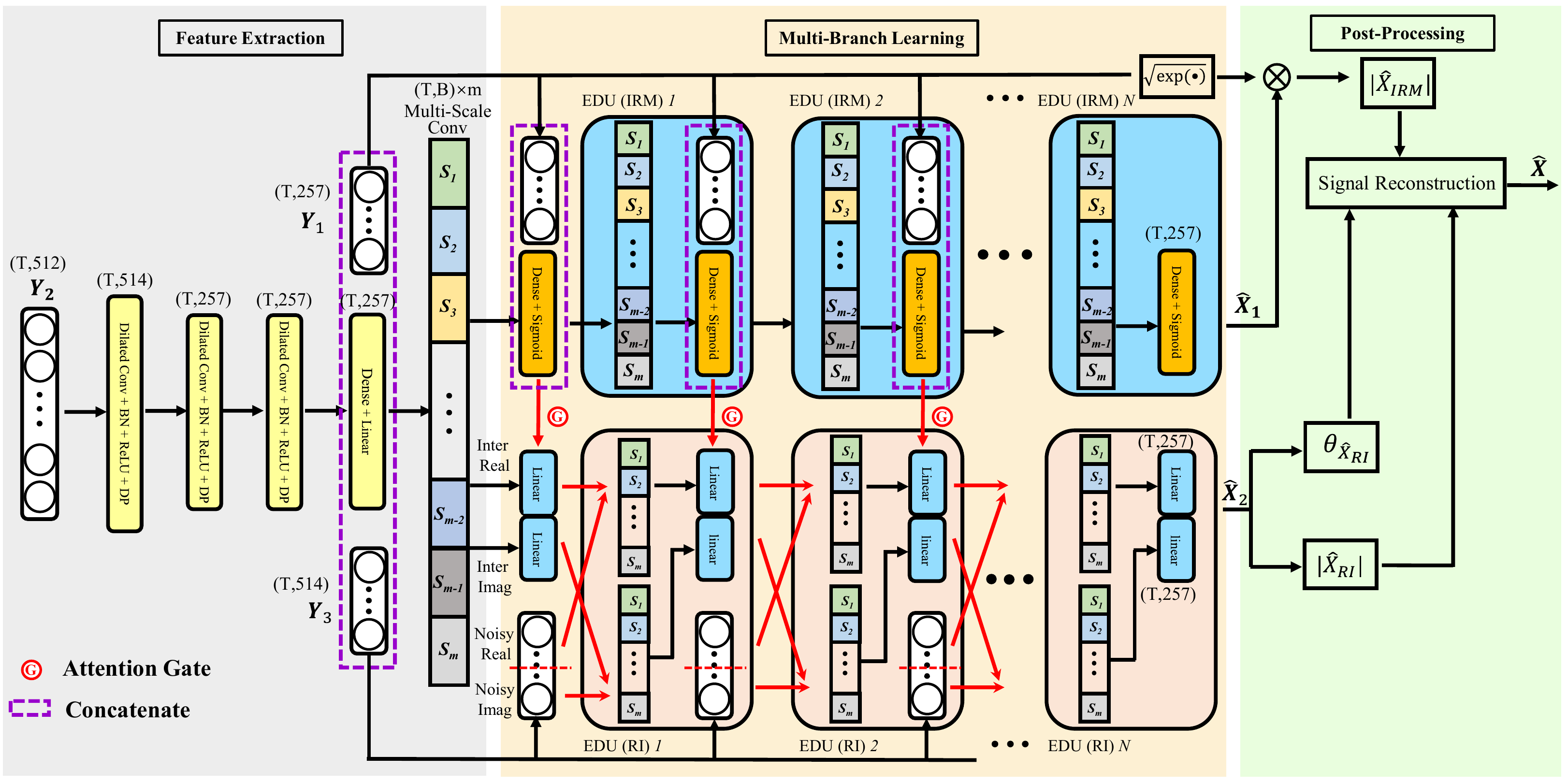}
\caption{Speech enhancement framework of the proposed PhaseDCN model.}
\label{fig:model}
\vspace{-0.1cm}
\end{figure*}

For TF-based speech enhancement models, the selection of input features plays an important role in improving noise reduction and model generalization \cite{chen2014feature}. Recently, dilated convolutions \cite{zhang2020multi,xu2020spex} are praised as a more effective temporal modeling approach to extract long-term acoustic features. In the feature extraction module of PhaseDCN, as shown in Figure~\ref{fig:model}, we superimpose layer by layer three dilated convolutional layers with a kernel size of 3 and dilated factor of 1, 3, and 5 respectively to capture the long-term features from the time domain signal \(Y_2\). The batch normalization (BN) \cite{ioffe2015batch}, ReLU activation and dropout (DP) \cite{srivastava2014dropout} are successively performed after each convolution. Then, a fully-connected layer (dense layer) is used to linearly transform the features. Besides, the local information of the current frame contributes the most to noise reduction. Therefore, the log power spectrum \(Y_1\) and the real and imaginary (RI) spectrum \(Y_3\) of the current frame are extracted and fused with the long-term features through the multi-scale dilated convolutional layer with a kernel size of 1 and a dilation factor of 1. Note that all these three kinds of input feature (\(Y_1\), \(Y_2\), and \(Y_3\)) are normalized to zero mean and unit variance, respectively, to speed up the training.

\subsection{Multi-Scale temporal analysis}

Considering the difference in speech speed of speakers, we propose a multi-scale convolutional method to achieve more fine-grained temporal feature analysis. As presented in Figure~\ref{fig:model}, this multi-scale temporal analysis is performed both in the feature extraction and multi-branch learning modules, and its definition is as follows:
\begin{equation}
  F\!_{m\!d\!,b\!}(\!t\!)\!=\!\left(\!Y\!\!*\!f\!_{m\!d\!,b\!}\right)\!=\!\sum_{i=0}^{K\!-\!1}\!f\!_{m\!d\!,b}(\!i\!)\!\cdot\!\left\{\!\tilde{F}\!_{m\!d\!,b\!-\!1\!},Y\!_{b}\!\right\}(\!t\!-\!d\!\cdot i)
  \label{eq1}
\end{equation}
Where \(f_{md,b}\) and \(F_{md,b}(t)\) are the multi-scale kernel and the output of sub-band, respectively. \(t\) is the frame index and \(T\) represents the number of frames. \(b\) is the sub-band index and its width is \(B\). \(Y_b\) represents the input features of each band, \(\tilde{F}_{md,b-1}\) is the output of the adjacent band corresponding to sub-band \(b\). \(K\) is the kernel size and \(d\) is the dilation factor, which determines the number of past frames for analysis.

As shown in Figure~\ref{fig:subfig:a}, the multi-scale temporal analysis performs sub-band convolutions from two directions. Each multi-scale layer is divided into \(m\) sub-bands, and the output of each sub-band considers the current input and the output of the previous sub-band. BN, ReLU, and DP are successively performed after each sub-band convolution. In this way, the receptive field of the sub-bands can increase linearly in the decomposition direction. The leftward multi-scale analysis is completed on the basis of the rightward results, which helps to balance the receptive field of each sub-band. Finally, the sub-band features in the two directions are spliced and added to obtain the multi-scale analysis result. It should be noted that in our PhaseDCN model, the dimensions of the multi-scale layers in the feature extraction module and the multi-branch learning module are 1028 and 514, respectively, and the numbers of sub-bands are 16 and 8, respectively.

\vspace{-0.2cm}
\subsection{Multi-branch learning architecture}
\vspace{-0.05cm}

In our PhaseDCN, we aim to separate the learning path of the IRM and complex spectrum, as the two targets may need different transformation characteristics. However, both paths use a common feature extraction module to extract features that are more universal and suitable for speech denoising tasks. In this multi-branch learning architecture, we adopt encoder-decoder units (EDUs) to achieve the progressive learning of the two targets. As shown in Figure~\ref{fig:model}, the input of EDU consists of two parts: an intermediate estimate of the target output and its corresponding original input. In EDU, the input features are firstly encoded by the multi-scale layer, and then decoded for the estimation of IRM or RI spectrum. In our model, we stack three EDUs with a kernel size of 3, using a dilated rate of 1, 3, and 5 respectively.

\begin{figure*}[t]
	\centering
		\begin{minipage}[l]{0.8\textwidth} %minipage使之保持同一行，0.2占这行的0.2
			\centering
			   \subfigure[]
			   {
                \label{fig:subfig:a}
                \includegraphics[width=0.63\textwidth]{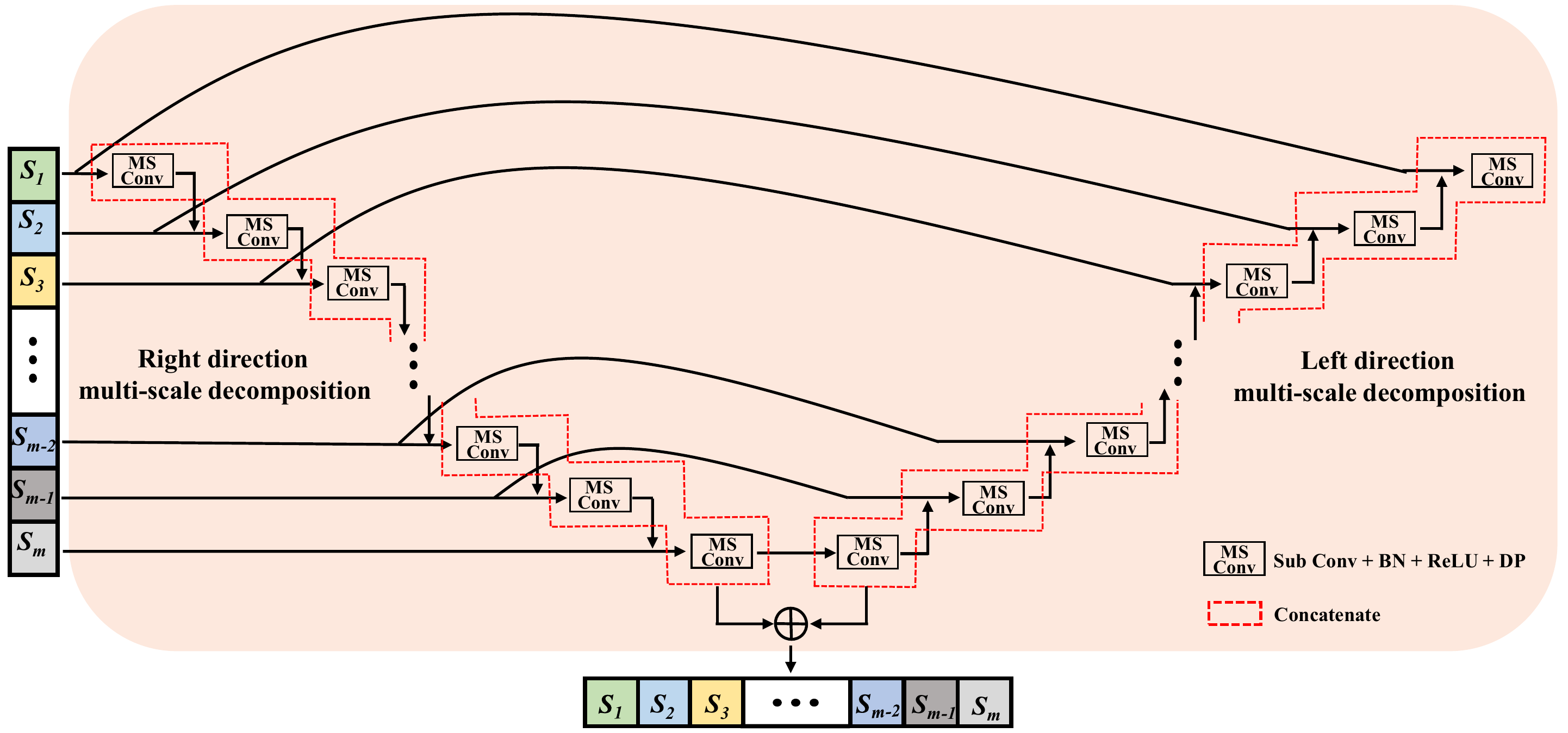}
                }
                \subfigure[]
                {
                \label{fig:subfig:b}
                \includegraphics[width=0.34\textwidth]{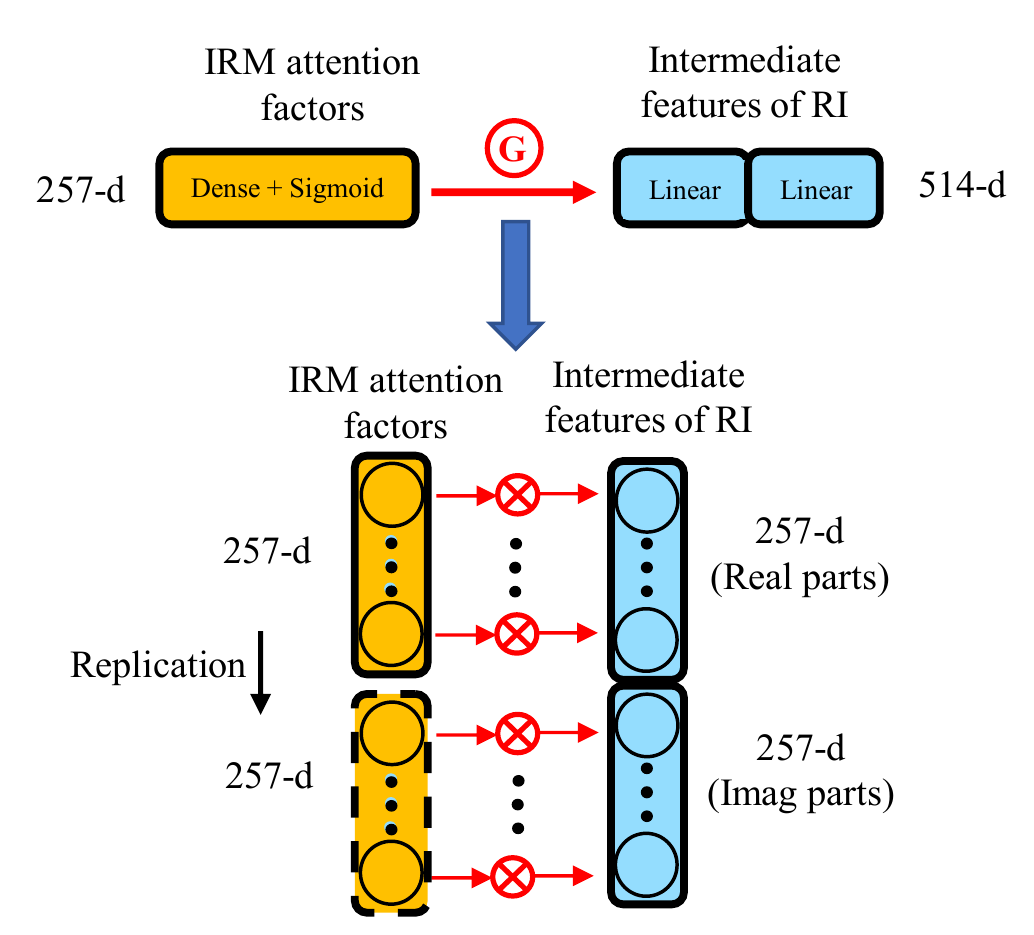}
                }
		\end{minipage}
	\caption{ (a) Principles of the multi-scale temporal analysis in both directions (b) IRM attention gate for multi-branch learning.}
	\label{fig:fig_micPerMon}
	\vspace{-0.2cm}
\end{figure*}

Meanwhile, since the IRM characterizes the probability of speech components in noisy signals to a certain extent, we introduce an attention control mechanism in the IRM path to achieve the information interaction between the two learning paths, as shown in Figure~\ref{fig:subfig:b}. The generated attention factors are duplicated and multiplied with the intermediate features of the RI spectrum to realize attention control on each TF point.

\subsection{Reconstruction for enhanced signals}

To better utilize the predicted outputs of the PhaseDCN model, we apply a post-processing way for signal reconstruction to achieve the complementary advantages of IRM and RI targets, as shown in Figure~\ref{fig:model}. For the output of the IRM path, the estimated IRM values are multiplied by the magnitude spectrum of the noisy speech to obtain the enhanced spectrum in a TF filtering way:
\begin{equation}
  \left|\hat{X}_{I\!R\!M}(k,t)\right|=\sqrt{\exp\left(Y_{1}(k,t)\right)}\cdot\hat{X}_{1}(k,t)
  \label{eq2}
\end{equation}
Where \(k\) and \(t\) are the frequency and frame index. For the output of the RI path, the reconstructed RI spectrum contributes to restoring both the enhanced magnitude and phase spectrum:
\begin{equation}
  \left|\hat{X}_{R\!I}(k,t)\right|=\sqrt{\operatorname{Re}\left(\hat{X}_{2}(k,t)\right)^{2}+\operatorname{Im}\left(\hat{X}_{2}(k,t)\right)^{2}}
  \label{eq3}
\end{equation}
\vspace{-0.3cm}
\begin{equation}
  \theta_{\hat{X}_{R\!I}}(k,t)=\arctan\left(\frac{\operatorname{Im}\left(\hat{X}_{2}(k, t)\right)}{\operatorname{Re}\left(\hat{X}_{2}(k,t)\right)}\right)
  \label{eq4}
\end{equation}
Then, we average the obtained two enhanced magnitude spectrum and reconstruct the signal using the enhanced phase as follows:
\vspace{-0.2cm}
\begin{equation}
  |\hat{X}(k,t)|=\frac{1}{2}\left(\left|\hat{X}_{R\!I}(k,t)\right|+\left|\hat{X}_{I\!R\!M}(k,t)\right|\right)
  \label{eq5}
\end{equation}
\vspace{-0.2cm}
\begin{equation}
 \hat{X}(k,t)=|\hat{X}(k,t)|\cdot\exp\left(i\theta_{\hat{X}_{R\!I}}(k,t)\right)
  \label{eq6}
\end{equation}
This not only reduces the impact of over and under estimation of the DNN predicted outputs, but also improves the phase distortion caused by background noises. Finally, the enhanced waveform is obtained through inverse STFT and overlap-add operations.

\vspace{-0.1cm}
\section{Experiment and Results}

\subsection{Experimental setups}

In our experiment, we evaluate the models on the TIMIT speech database \cite{garofolo1988getting} corrupted by the noises from NOISEX-92 database \cite{varga1993assessment}. A 38-hour training dataset is constructed by mixing 4,620 utterances from the TIMIT training set with 12 noises (\textit{babble, factory1, destroyer1, destroyer2, cockpit1, cockpit2, volvo, tank, leopard, white, hfchannel, machinegun}) from NOISEX-92. Each utterance is mixed with the first 60$\%$ part of each noise file, and the mixed SNR level follows the uniform distribution in the range of -5 to 15. Similarly, 280 utterances from the TIMIT test set are mixed with the middle 20$\%$ of each noise to produce a validation dataset for model training. To evaluate the speaker generalization ability of the DNN models, 320 untrained utterances from the TIMIT test set are mixed with the last 20$\%$ of each noise as the seen noise test set. Besides, 3 new noises (\textit{cockpit3, factory2, pink}) from NOISEX-92 are mixed with the 320 utterances as the unseen noise test set to evaluate the noise generalization. The test SNR levels are fixed at \{-5, 0, 5, 10, 15\} dBs. We adopt short-time objective intelligibility (STOI) \cite{taal2011algorithm} and perceptual evaluation of speech quality (PESQ) \cite{rix2001perceptual} as two evaluation metrics.

All the utterances are resampled to 16 kHz, the frames are analyzed by a Hamming window with 32 ms length and 16ms overlap. The input feature lengths of LPS, waveform, and RI spectrum are 257, 512, and 514, respectively. The proposed models are trained using the equally weighted mean squared error (MSE) loss on the two targets. The Adam \cite{kingma2014adam} is used to optimize the models on every mini-batch with a batch size of 10,000 consecutive input frames. The dropout rate in our PhaseDCN model is set to 0.2.
%Aliquam quis orci consectetur nulla luctus ullamcorper. Suspendisse finibus luctus erat a dapibus.

\subsection{Ablation study for different model components}

In this section, an ablation study is conducted to evaluate the effectiveness of different model components of PhaseDCN. As presented in Table~\ref{tab:averaged results}, we compare the contributions of the multi-branch structure, multi-scale layer (denoted as ‘MS’), and IRM attention mechanism (denoted as ‘A’).

\begin{table}[th]
  \caption{The averaged PESQ and STOI results for different model configurations in seen and unseen noise cases}
  \label{tab:averaged results}
  \centering
  \resizebox{0.96\columnwidth}{!}{
  \begin{tabular}{c|cc|cc}
    \toprule
    \multirow{2}{*}{Methods} & \multicolumn{2}{c|}{\textbf{PESQ}} & \multicolumn{2}{c}{\textbf{STOI}} \\
                         & Seen           & Unseen           & Seen           & Unseen           \\
    \midrule
    Noisy                & 2.18           & 2.05             & 0.794          & 0.781             \\
    IRM-Branch-MS        & 3.02           & 2.73             & 0.891          & 0.863             \\
    RI-Branch-MS         & 2.83           & 2.65             & 0.887          & 0.870             \\
    Multi-Branch-MS      & 3.10           & 2.81             & 0.908          & 0.884              \\
    Multi-Branch-A       & 3.15           & 2.83             & 0.910          & 0.886              \\
    Multi-Branch-MS-A    & \textbf{3.17}  & \textbf{2.86}    & \textbf{0.915} & \textbf{0.890}      \\
    \bottomrule
  \end{tabular}
  }
\end{table}

\begin{table*}[th]
  \centering
  \caption*{{\rm Table 3: }Model comparisons in terms of STOI, PESQ, model size and FLOPs (counted in millions)}
  \label{tab:Model comparisons}
  \centering
  \resizebox{2.1\columnwidth}{!}
  {
  \begin{tabular}{cc|cccccc|cccccc|cc}
    \toprule
    \multicolumn{2}{c|}{\textbf{Metrics}} & \multicolumn{6}{c|}{\textbf{PESQ}} & \multicolumn{6}{c|}{\textbf{STOI}} & \textbf{Model}  & \multirow{2}{*}{\textbf{FLOPs}}    \\
    \multicolumn{1}{c|}{\textbf{Noises}}&\textbf{Methods}&\textbf{-5dB}&\textbf{0dB}&\textbf{5dB}&\textbf{10dB}&\textbf{15dB}&\textbf{Ave}&\textbf{-5dB}&\textbf{0dB}&\textbf{5dB}&\textbf{10dB}&\textbf{15dB}&\textbf{Ave}&\textbf{Size}   \\
    \midrule
    \multicolumn{1}{c|}{\multirow{5}{*}{\textbf{Seen}}}&Noisy  &1.49&1.82&2.18&2.53&2.87&2.18&0.623&0.718&0.809&0.884&0.937&0.794&--&-- \\
            \multicolumn{1}{c|}{} &MS-TCN       &2.45&2.80&3.10&3.35&3.58&3.06&0.791&0.863&0.911&0.943&0.964&0.894&7.7M&15.6M \\
            \multicolumn{1}{c|}{} &TCNN         &2.21&2.54&2.78&2.97&3.11&2.72&0.771&0.855&0.903&0.930&0.945&0.881&6.3M&12.5M \\
            \multicolumn{1}{c|}{} &GCRN         &2.41&2.87&3.11&3.36&3.53&3.06&0.810& \textbf{0.892}& \textbf{0.934}&0.957&0.969&0.912&9.8M&47.7M \\
            \multicolumn{1}{c|}{} &PhaseDCN     & \textbf{2.59}& \textbf{2.94}& \textbf{3.22}& \textbf{3.45}& \textbf{3.66}& \textbf{3.17}& \textbf{0.816}&0.890&0.933& \textbf{0.959}& \textbf{0.975}& \textbf{0.915}&7.5M&15.1M \\
    \midrule
    \multicolumn{1}{c|}{\multirow{5}{*}{\textbf{Unseen}}}&Noisy         &1.32&1.67&2.05&2.42&2.79&2.05&0.583&0.698&0.802&0.884&0.939&0.781&--&-- \\
            \multicolumn{1}{c|}{} &MS-TCN       &2.08&2.50&2.84&3.14&3.39&2.79&0.727&0.829&0.894&0.936&0.961&0.869&7.7M&15.6M \\
            \multicolumn{1}{c|}{} &TCNN         &2.03&2.39&2.68&2.91&3.08&2.62&0.742&0.844&0.902&0.932&0.946&0.873&6.3M&12.5M \\
            \multicolumn{1}{c|}{}&GCRN         &2.03&2.51&2.89& \textbf{3.22}& \textbf{3.46}&2.82& \textbf{0.755}& \textbf{0.868}& \textbf{0.920}&0.950&0.967 &\textbf{0.892}&9.8M&47.7M \\
            \multicolumn{1}{c|}{} &PhaseDCN     & \textbf{2.19}& \textbf{2.60}& \textbf{2.90}&3.18&3.44& \textbf{2.86}&0.753&0.857&0.916& \textbf{0.951}& \textbf{0.972}&0.890&7.5M&15.1M \\

    \bottomrule \\
  \end{tabular}
  }
  \vspace{-0.5cm}
\end{table*}

It can be seen from the first three comparison results that the performance of the multi-branch structure is obviously better than that of the other two single-branch models in seen and unseen noise cases. In addition, the comparison results between the ‘Multi-Branch-MS-A’ and the ‘Multi-Branch-MS’ indicate that the introduced IRM attention mechanism is an effective method to improve the enhanced speech quality and intelligibility. From the last two comparative experiments, the introduction of multi-scale layers helps to further improve the speech denoising effect. Besides, it can save 36$\%$ trainable parameters and reduce the floating-point of operations (FLOPs) per frame from 23.3 M to 15.1 M, making the model more efficient for inference.

\subsection{The evaluation of post-processing module}

The contributions of phase enhancement and different signal synthesis methods in our PhaseDCN are evaluated in Table 2. Among them, the abbreviations ‘IRM’ and ‘RI’ respectively represent the reconstruction of enhanced magnitude using the predicted IRM and RI spectrum, and ‘Ave’ represents the average of the two targets in the magnitude domain. ‘UnPha’ and ‘EnPha’ represent signal synthesis using the noisy phase and the phase of enhanced RI spectrum respectively.

\begin{table}[th]
  \caption{The averaged PESQ and STOI results for different signal reconstruction methods in seen and unseen noise cases}
  \label{tab:averaged reconstruction results}
  \centering
  \resizebox{0.83\columnwidth}{!}{
  \begin{tabular}{c|cc|cc}
    \toprule
    \multirow{2}{*}{Methods} & \multicolumn{2}{c|}{\textbf{PESQ}} & \multicolumn{2}{c}{\textbf{STOI}} \\
                         & Seen           & Unseen           & Seen           & Unseen           \\
    \midrule
    Noisy&2.18&2.05&0.794&0.781\\
    IRM-UnPha&3.04&2.74&0.893&0.866\\
    IRM-EnPha&3.08&2.78&0.905&0.878\\
    RI-EnPha&3.06&2.79&0.906&0.886\\
    Ave-UnPha&3.13&2.82&0.898&0.873\\
    Ave-EnPha& \textbf{3.17}& \textbf{2.86}& \textbf{0.915}& \textbf{0.890}\\
    \bottomrule
  \end{tabular}
  }
\end{table}

From the above Table~\ref{tab:averaged reconstruction results}, we can find that the enhanced phase brings improvements in PESQ and STOI results for both synthesis methods using a single target and two targets. The reduction of phase distortion contributes more improvement to speech intelligibility. Compared with the ‘Ave-UnPha’, the STOI of ‘Ave-EnPha’ is increased by 1.7$\%$ by only replacing the noisy phase with the enhanced phase. Table~\ref{tab:averaged reconstruction results} also indicates that the proposed ‘Ave-EnPha’ synthesis method achieves the complementary advantages of masking and mapping targets, resulting in a more comprehensive noise reduction effect of noisy speech in both speech quality and intelligibility.

\subsection{Comparison with other advanced models}

In this section, we compare PhaseDCN with three advanced DNN-based causal speech enhancement models. MS-TCN [17] is our previous multi-scale temporal convolutional network (TCN) model used for the multi-objective learning of IRM and LPS. TCNN [16] and GCRN [13] are two state-of-the-art phase-aware speech enhancement methods, which carry out the waveform and complex spectral mapping, respectively. For the TCNN model, in order to keep the sequence length unchanged, we use the same convolution instead of the valid convolution in the encoder and decoder parts, and other configurations are consistent with the original. For the GCRN model, we use its best configuration (the number of groups is 2) for comparison.

Table 3 reports the evaluated PESQ and STOI results of different models and their model sizes and FLOPs. The proposed PhaseDCN outperforms the MS-TCN in both seen and unseen noise cases. Although MS-TCN also optimizes the masking and mapping targets in its output, the deep interactive learning way of PhaseDCN can better predict two targets. Moreover, the introduction of the RI spectrum enables PhaseDCN to effectively reduce phase distortion, thereby obtaining better speech quality and intelligibility. Among the reported phase-aware methods, the PhaseDCN and GCRN perform better, and their performance is significantly better than the TCNN model. In contrast with GCRN, the PhaseDCN model achieves better PESQ and similar STOI results with only 77$\%$ parameters and 32$\%$ FLOPs. Furthermore, the PhaseDCN model shows more significant advantages in the case of low SNR (-5 dB). As a result, PhaseDCN is more cost-effective in terms of performance and complexity for low-latency speech enhancement tasks.

\vspace{-0.15cm}
\section{Conclusions}

In this paper, we introduce PhaseDCN, a multi-branch dilated convolution network, to aggregate multi-scale context in IRM and RI spectrum for the single-channel speech enhancement task. Learning the RI spectrum effectively reduces the phase distortion caused by background noises in low SNRs. The incorporation of the IRM attention mechanism is helpful to improve multi-objective learning. Stacking multi-scale dilated convolutions not only enlarges the receptive field of PhaseDCN at a more granular level, but also significantly elevates the inference speed of the model. Due to its causality and excellent noise reduction effect, the proposed PhaseDCN is more suitable for some low-latency human hearing applications.

\section{Acknowledgements}

This work was supported in part by the Basic Research Program under Grants No. JCYJ20170412151226061 and No. JCYJ20180507182241622 funded by Shenzhen government.

\bibliographystyle{IEEEtran}

\bibliography{mybib}

\begin{thebibliography}{10}
\providecommand{\url}[1]{#1}
\csname url@samestyle\endcsname
\providecommand{\newblock}{\relax}
\providecommand{\bibinfo}[2]{#2}
\providecommand{\BIBentrySTDinterwordspacing}{\spaceskip=0pt\relax}
\providecommand{\BIBentryALTinterwordstretchfactor}{4}
\providecommand{\BIBentryALTinterwordspacing}{\spaceskip=\fontdimen2\font plus
\BIBentryALTinterwordstretchfactor\fontdimen3\font minus
  \fontdimen4\font\relax}
\providecommand{\BIBforeignlanguage}[2]{{%
\expandafter\ifx\csname l@#1\endcsname\relax
\typeout{** WARNING: IEEEtran.bst: No hyphenation pattern has been}%
\typeout{** loaded for the language `#1'. Using the pattern for}%
\typeout{** the default language instead.}%
\else
\language=\csname l@#1\endcsname
\fi
#2}}
\providecommand{\BIBdecl}{\relax}
\BIBdecl

\bibitem{scalart1996speech}
P.~Scalart \emph{et~al.}, ``Speech enhancement based on a priori signal to
  noise estimation,'' in \emph{ICASSP}, 1996, pp. 629--632.

\bibitem{ephraim1985speech}
Y.~Ephraim and D.~Malah, ``Speech enhancement using a minimum mean-square error
  log-spectral amplitude estimator,'' \emph{IEEE Transactions on Acoustics,
  Speech, and Signal processing}, vol.~33, no.~2, pp. 443--445, 1985.

\bibitem{cohen2001speech}
I.~Cohen and B.~Berdugo, ``Speech enhancement for non-stationary noise
  environments,'' \emph{Signal Processing}, vol.~81, no.~11, pp. 2403--2418,
  2001.

\bibitem{xu2014regression}
Y.~Xu, J.~Du, L.-R. Dai, and C.-H. Lee, ``A regression approach to speech
  enhancement based on deep neural networks,'' \emph{IEEE/ACM Transactions on
  Audio, Speech, and Language Processing}, vol.~23, no.~1, pp. 7--19, 2014.

\bibitem{kumar2016speech}
A.~Kumar and D.~Florencio, ``Speech enhancement in multiple-noise conditions
  using deep neural networks,'' in \emph{INTERSPEECH}, 2016, pp. 3738--3742.

\bibitem{park2016fully}
S.~R. Park and J.~Lee, ``A fully convolutional neural network for speech
  enhancement,'' in \emph{INTERSPEECH}, 2017, pp. 1993--1997.

\bibitem{tan2018gated}
K.~Tan, J.~Chen, and D.~Wang, ``Gated residual networks with dilated
  convolutions for monaural speech enhancement,'' \emph{IEEE/ACM Transactions
  on Audio, Speech, and Language Processing}, vol.~27, no.~1, pp. 189--198,
  2018.

\bibitem{gerkmann2015phase}
T.~Gerkmann, M.~Krawczyk-Becker, and J.~Le~Roux, ``Phase processing for
  single-channel speech enhancement: History and recent advances,'' \emph{IEEE
  Signal Processing Magazine}, vol.~32, no.~2, pp. 55--66, 2015.

\bibitem{williamson2015complex}
D.~S. Williamson, Y.~Wang, and D.~Wang, ``Complex ratio masking for monaural
  speech separation,'' \emph{IEEE/ACM Transactions on Audio, Speech, and
  Language Processing}, vol.~24, no.~3, pp. 483--492, 2015.

\bibitem{wang2019masking}
X.~Wang and C.~Bao, ``Masking estimation with phase restoration of clean speech
  for monaural speech enhancement.'' in \emph{INTERSPEECH}, 2019, pp.
  3188--3192.

\bibitem{hu2020dccrn}
Y.~Hu, Y.~Liu, S.~Lv, M.~Xing, S.~Zhang, Y.~Fu, J.~Wu, B.~Zhang, and L.~Xie,
  ``{DCCRN}: Deep complex convolution recurrent network for phase-aware speech
  enhancement,'' in \emph{INTERSPEECH}, 2020, pp. 2472--2476.

\bibitem{li2020two}
A.~Li, C.~Zheng, R.~Peng, and X.~Li, ``Two heads are better than one: A
  two-stage approach for monaural noise reduction in the complex domain,''
  \emph{arXiv preprint arXiv:2011.01561}, 2020.

\bibitem{tan2019learning}
K.~Tan and D.~Wang, ``Learning complex spectral mapping with gated
  convolutional recurrent networks for monaural speech enhancement,''
  \emph{IEEE/ACM Transactions on Audio, Speech, and Language Processing},
  vol.~28, pp. 380--390, 2019.

\bibitem{pascual2017segan}
S.~Pascual, A.~Bonafonte, and J.~Serra, ``{SEGAN}: Speech enhancement
  generative adversarial network,'' in \emph{INTERSPEECH}, 2017, pp.
  3642--3646.

\bibitem{fu2018end}
S.-W. Fu, T.-W. Wang, Y.~Tsao, X.~Lu, and H.~Kawai, ``End-to-end waveform
  utterance enhancement for direct evaluation metrics optimization by fully
  convolutional neural networks,'' \emph{IEEE/ACM Transactions on Audio,
  Speech, and Language Processing}, vol.~26, no.~9, pp. 1570--1584, 2018.

\bibitem{pandey2019tcnn}
A.~Pandey and D.~Wang, ``{TCNN}: Temporal convolutional neural network for
  real-time speech enhancement in the time domain,'' in \emph{ICASSP}, 2019,
  pp. 6875--6879.

\bibitem{zhang2020multi}
L.~Zhang and M.~Wang, ``{Multi-Scale TCN}: Exploring better temporal {DNN}
  model for causal speech enhancement,'' in \emph{INTERSPEECH}, 2020, pp.
  2672--2676.

\bibitem{li2020recursive}
A.~Li, C.~Zheng, C.~Fan, R.~Peng, and X.~Li, ``A recursive network with dynamic
  attention for monaural speech enhancement,'' in \emph{INTERSPEECH}, 2020, pp.
  2422--2426.

\bibitem{chen2014feature}
J.~Chen, Y.~Wang, and D.~Wang, ``A feature study for classification-based
  speech separation at low signal-to-noise ratios,'' \emph{IEEE/ACM
  Transactions on Audio, Speech, and Language Processing}, vol.~22, no.~12, pp.
  1993--2002, 2014.

\bibitem{xu2020spex}
C.~Xu, W.~Rao, E.~S. Chng, and H.~Li, ``{SpEx}: Multi-scale time domain speaker
  extraction network,'' \emph{IEEE/ACM Transactions on Audio, Speech, and
  Language Processing}, vol.~28, pp. 1370--1384, 2020.

\bibitem{ioffe2015batch}
S.~Ioffe and C.~Szegedy, ``Batch normalization: Accelerating deep network
  training by reducing internal covariate shift,'' in \emph{ICML}, 2015, pp.
  448--456.

\bibitem{srivastava2014dropout}
N.~Srivastava, G.~Hinton, A.~Krizhevsky, I.~Sutskever, and R.~Salakhutdinov,
  ``Dropout: a simple way to prevent neural networks from overfitting,''
  \emph{The Journal of Machine Learning Research}, vol.~15, no.~1, pp.
  1929--1958, 2014.

\bibitem{garofolo1988getting}
J.~S. Garofolo, L.~F. Lamel, W.~M. Fisher, J.~G. Fiscus, and D.~S. Pallett,
  ``Getting started with the {DARPA TIMIT CD-ROM}: An acoustic phonetic
  continuous speech database,'' \emph{National Institute of Standards and
  Technology (NIST), Gaithersburgh, MD}, 1988.

\bibitem{varga1993assessment}
A.~Varga and H.~J. Steeneken, ``Assessment for automatic speech recognition:
  {II. NOISEX-92}: A database and an experiment to study the effect of additive
  noise on speech recognition systems,'' \emph{Speech Communication}, vol.~12,
  no.~3, pp. 247--251, 1993.

\bibitem{taal2011algorithm}
C.~H. Taal, R.~C. Hendriks, R.~Heusdens, and J.~Jensen, ``An algorithm for
  intelligibility prediction of time--frequency weighted noisy speech,''
  \emph{IEEE Transactions on Audio, Speech, and Language Processing}, vol.~19,
  no.~7, pp. 2125--2136, 2011.

\bibitem{rix2001perceptual}
A.~W. Rix, J.~G. Beerends, M.~P. Hollier, and A.~P. Hekstra, ``Perceptual
  evaluation of speech quality ({PESQ})-a new method for speech quality
  assessment of telephone networks and codecs,'' in \emph{ICASSP}, 2001, pp.
  749--752.

\bibitem{kingma2014adam}
D.~P. Kingma and J.~Ba, ``Adam: A method for stochastic optimization,'' in
  \emph{ICLR}, 2014, pp. 1--13.

\end{thebibliography}

% \begin{thebibliography}{9}
% \bibitem[1]{Davis80-COP}
%   S.\ B.\ Davis and P.\ Mermelstein,
%   ``Comparison of parametric representation for monosyllabic word recognition in continuously spoken sentences,''
%   \textit{IEEE Transactions on Acoustics, Speech and Signal Processing}, vol.~28, no.~4, pp.~357--366, 1980.
% \bibitem[2]{Rabiner89-ATO}
%   L.\ R.\ Rabiner,
%   ``A tutorial on hidden Markov models and selected applications in speech recognition,''
%   \textit{Proceedings of the IEEE}, vol.~77, no.~2, pp.~257-286, 1989.
% \bibitem[3]{Hastie09-TEO}
%   T.\ Hastie, R.\ Tibshirani, and J.\ Friedman,
%   \textit{The Elements of Statistical Learning -- Data Mining, Inference, and Prediction}.
%   New York: Springer, 2009.
% \bibitem[4]{YourName17-XXX}
%   F.\ Lastname1, F.\ Lastname2, and F.\ Lastname3,
%   ``Title of your INTERSPEECH 2021 publication,''
%   in \textit{Interspeech 2021 -- 20\textsuperscript{th} Annual Conference of the International Speech Communication Association, September 15-19, Graz, Austria, Proceedings, Proceedings}, 2020, pp.~100--104.
% \end{thebibliography}

\end{document}